\documentclass{aa}
\usepackage[dvips]{graphicx}
\def\simless{\mathbin{\lower 3pt\hbox
{$\rlap{\raise 5pt\hbox{$\char'074$}}\mathchar"7218$}}}   
\def\simmore{\mathbin{\lower 3pt\hbox
{$\rlap{\raise 5pt\hbox{$\char'076$}}\mathchar"7218$}}}   
\def\Msun{{\rm M}_\odot}                                       
\newcommand{\be}{\begin{equation}}
\newcommand{\ee}{\end{equation}}

\begin{document}

\title{Powerful GeV emission from a $\gamma$-ray-burst shock wave 
scattering stellar photons}
\titlerunning{Powerful GeV emission from a $\gamma$-ray-burst shock wave 
scattering stellar photons}
\authorrunning{Dimitrios Giannios}
\author{Dimitrios Giannios}

\institute{Max Planck Institute for Astrophysics, Box 1317, D-85741 Garching, Germany}

\offprints{giannios@mpa-garching.mpg.de}
\date{Received / Accepted}

\abstract
{The $\gamma$-ray bursts (GRBs) of long duration are very likely to be connected to the
death of massive stars. The $\gamma$-ray emission is believed to come from
energy released internally in a flow that moves at ultrarelativistic speed. 
The fast flow drives a shock wave into the external medium leading to 
the afterglow emission. Most massive stars form in dense clusters, their
high luminosity producing a very dense radiation field. Here, I explore
the observational consequences of the interaction of the shocked external
medium of the burst with the photon field of a nearby O star. I show
that inverse Compton scattering of the stellar 
photons by electrons heated by the shock leads to powerful $\gamma$-ray emission  
in the $\sim 1-100$ GeV range. This emission appears minutes to hours after
the burst and can be easily detected by Cherenkov telescopes and 
probably with the GLAST satellite. This signal may have already been observed
in GRB 940217 and can yield important information about the circumburst
environment and the extragalactic background light.
\\   
{\bf Key words:} Gamma rays: bursts -- radiation mechanisms: general}

\maketitle

\section{Introduction} 
\label{intro}

Long-duration $\gamma$-ray bursts (GRBs) are believed to result from the
death of Wolf-Rayet stars (MacFadyen \& woosley 1999; Galama et al. 1998; Stanek
et al. 2003; Hjorth et al. 2003). Stars rarely form in isolation. This is particularly true 
for the most massive stars, a large fraction of which form in high-mass clusters
of $M_{\rm c}\simmore 10^4 \Msun$. About 30\% of the known Galactic
Wolf-Rayet stars are located in the high-mass clusters
Westerlund 1 and the Galactic center clusters (Arches, Quintuplet, and 
Center; e.g. see Figer 2004; Crowther 2007).     

High-mass clusters are characterized by a high concentration of
O stars with luminosity about a million times the solar. 
Such a cluster contains tens to hundreds of O-stars in a 
typical radius of $3\times 10^{17}-1\times 10^{18}$ cm, so the
mean distance between them is $\sim 1-3\times 10^{17}$ cm (Massey \&
Hunter 1998; Figer 2004). A large fraction of GRBs is, therefore, expected 
to take place in a dense environment of luminous stars. 

The GRB emission is believed to be the result of internal energy dissipation
in a relativistic flow. Additional X-ray/$\gamma$-ray emission is
expected as the result of inverse Compton scattering if the fast flow transverses a 
dense radiation field. This can take place in the funnel of the collapsing
star (Lazzati et al. 2000), during the jet breakout from the collapsar (MacFadyen et
al. 2001) or shortly afterwards because of the presence of a stellar companion (Ramirez-Ruiz 2004). 

At a much larger distance from the central engine,
the ejecta decelerate, interacting with the external medium driving 
a shock into it. For typical GRB parameters the
blast wave remains relativistic at $\sim 0.1$pc distance where it 
can interact with a nearby O star. In this paper, I study the $\gamma$-ray 
emission from this interaction.
The O star provides the UV photons that are upscattered (external
Compton) by the electrons accelerated in the forward shock. 
It is shown that such interaction leads to powerful afterglow emission 
in the $\sim 1-100$ GeV energy range.

\section{The model}

Consider a Wolf-Rayet star exploding at a distance  $R_{\rm s}\simmore
 10^{17}$ cm from an O star and launching the ultrarelativistic
outflow (with bulk $\Gamma\simmore 100$) that produces
a $\gamma$-ray burst. After the burst has taken place, the flow interacts 
with the external medium driving a relativistic shock into it.
The afterglows of GRBs seen in X rays and lower frequencies are often
interpreted as synchrotron emission from this shock (Sari et al. 1998). 
When the shock approaches the O star, it
sweeps through its dense  radiation field. Inverse Compton scattering of 
this radiation field is an effective way of converting the energy of the shocked
medium into high-energy photons.

\subsection{Inverse Compton in the forward shock}

The expected observable flux and energy of the photons 
of this emission is determined by the energy density of the 
stellar photon field and the number and energy of the scattering 
particles in the blast wave.
 
The shock propagates into a circumburst medium with a  
complex density profile that is shaped by the interacting 
winds of the two stars. For typical values of the mass loss rates and wind speeds 
of the Wolf-Rayet and O stars (Stevens et al. 1992; Chevalier \& Li 2000) and for
a distance $R_{\rm s}\sim$ a few $10^{17}$ cm, the particle densities 
of the medium are of order $n\sim 10^2$ cm$^{-3}$. 

The blast wave is still ultrarelativistic at such distances 
from the source. For a spherical and adiabatic shock, the shocked 
medium at distance $R_{\rm s}$ has a Lorentz factor of 
(Blandford \& McKee 1976)
\be
\Gamma\simeq \Big(\frac{3E}{4\pi R_{\rm s}^3n m_{\rm p}
  c^2}\Big)^{1/2}\simeq 40\frac{E_{54}^{1/2}}{n_2^{1/2}R_{17}^{3/2}},
\label{Gamma} 
\ee 
 where $m_{\rm p}$ and $c$ are the proton mass and the speed of
light, and the notation $A=10^xA_x$ and CGS units are used. Also, $E$
is the total energy of the shocked medium (kinetic and thermal). 
The value $10^{54}$ erg is typical of the equivalent isotropic energy 
inferred for GRBs. An estimated fraction of some 10\% of this energy appears 
as burst emission, but most of the energy (depending on model parameters) converts into an outward
propagating shock, and is available for the afterglow process proposed here.

If the energy that is dissipated in the shock were efficiently thermalized 
and shared among the particles, electrons would be accelerated to a characteristic
Lorentz factor $\gamma_{\rm e}\sim \Gamma m_{\rm p}/m_{\rm e}$ in the 
frame comoving with the shocked medium. However, in the shocked medium under
consideration, Coulomb collisions are not an efficient channel for energy
exchange between particles and the fate of the energy released in the shock is not clear.   
I assume, as is customary in afterglow models, that the electrons swept by the shock 
receive a fraction $\epsilon_{\rm e}\sim 0.1$ of the dissipated energy. 
In the frame comoving with the blast wave, the electron distribution
is assumed isotropic with a characteristic Lorentz factor (Sari et al. 1998;
Wijers \& Galama 1999)
\be
\gamma_{\rm e}=\epsilon_{\rm e}\Gamma m_{\rm p}/m_{\rm e}\simeq 7.4 \cdot10^3
\frac{\epsilon_{\rm e,-1}E_{54}^{1/2}} {n_2^{1/2}R_{17}^{3/2}},
\ee
where  $\epsilon_{\rm e}= 0.1\epsilon_{\rm e,-1}$.

\subsubsection{Energy of the inverse Compton photons}
In the frame of the collapsing star the energy of the heated electrons is 
$E_{\rm e}=4\Gamma \gamma_{\rm e} m_{\rm e} c^2/3$. They transfer part of 
this energy to the stellar photons through
scattering. The scattered photons gain a factor $A\sim 2\Gamma^2\gamma_{\rm e}^2$
with respect to their initial energy $e_{\rm s}$.  Including recoil, the
energy gain of a photon is limited to the energy of the scattering electron
$E_{\rm e}$. The characteristic energy of the 
{\it observed} inverse Compton emission for a burst at redshift $z$ is then
\be
E_\gamma={\rm min}\Big[\frac{1700}{1+z}
 \frac{ \epsilon_{\rm e,-1}^2E_{54}^{2}}{n_2^2R_{17}^6} e_{s,1}, \frac{200}{1+z}
 \frac{ \epsilon_{\rm e,-1}E_{54}}{n_2R_{17}^3}\Big]\quad {\rm GeV},
\label{Egamma} 
\ee
where  $e_{\rm s}=10e_{\rm s,1}$ eV.
The energy of the typical stellar photon is $e_{\rm s}\simeq 2.7k_{\rm
B}T_{\rm s}\simeq 10$ eV for a main-sequence O star of temperature $T_{\rm s}\sim$
45,000 K. If the star has evolved into a red supergiant, its temperature is 
a factor of 10 less, resulting in $e_{\rm s}\sim 1$ eV.

\subsubsection{Observed photon fluence}
Since the blast wave moves with a Lorentz factor $\Gamma\gg 1$, 
the observer sees only emission coming from a cone of opening angle
$\sim 1/\Gamma$ with respect to the line of sight.    
The flux of high-energy photons expected from inverse Compton
scattering depends on the distance of the O star from the line
of sight to the GRB. Let $\theta$ be the angle between the O star
and the observer, as seen from the GRB source, and then distinguish between the
{\it on-axis} case  where $\theta\sim 1/\Gamma$ and the {\it off-axis} case 
in which $\theta\sim 1$.

\begin{figure}
\resizebox{\hsize}{!}{\includegraphics[]{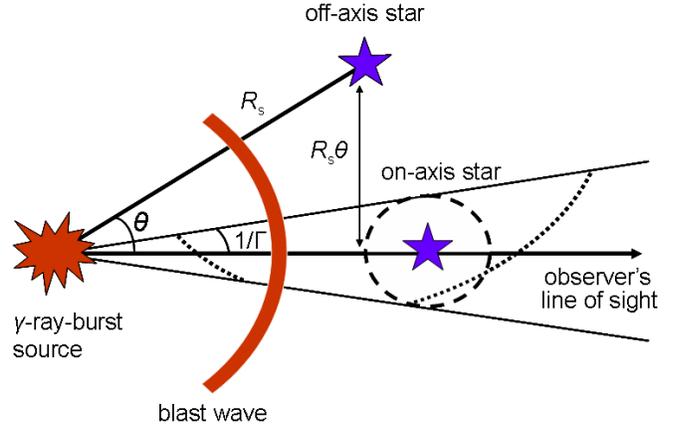}}
\caption[] { High-energy photons produced by a relativistic 
blast wave propagating with a Lorentz factor $\Gamma$ through the radiation 
field of a luminous O star, located at a distance $R_{\rm s}$ from the center 
of the explosion and an angle $\theta$ with respect to the line of sight to 
the observer. The observer sees only scattered photons that lie within a cone
of opening angle $\sim 1/\Gamma$ ({\it observer's cone}). 
In the {\it on-axis} case, the O star is 
inside the observer's cone. An {\it off-axis} star is located at $\theta\sim 1$. 
The dashed and dotted curves mark the region where most of the
observed scattering events take place for on-axis and off-axis stars
respectively. 
\label{fig1}}
\end{figure}

The shocked interstellar material forms a thin shell of width $\sim R/
\Gamma^2$ behind the relativistic blast wave (Blandford \& McKee 1976). 
When it reaches distance $R_{\rm s}$, the blast wave has swept through $N_{\rm
t}=4\pi nR_{\rm s}^3/3$ electrons out of which $N=N_{\rm
  t}/4\Gamma^2 =\pi nR_{\rm s}^3/3\Gamma^2$ are radiating within the observer's cone. These
shock-heated electrons have a mean Lorentz factor $4\gamma_{\rm e}\Gamma/3$
(in the frame of the collapsing star). Using the standard
expression for Compton scattering by relativistic electrons, the  instantaneous energy
flux of the up-scattered photons is $F(t)=4c\sigma_{\rm T} U_{\rm rad}
(4\gamma_{\rm e}\Gamma/3)^2 N/3$. The O star's radiation field $U_{\rm rad}$ can be
approximated by its average $\bar{U}_{\rm rad}$ over the volume V where most scatterings take
place (see Fig.~1). 

Integrating the resulting flux in time as the shell crosses through V  
gives the expression 
\be
F_\gamma\sim \frac{\Gamma L_{\rm s}\sigma_{\rm T}nR_{\rm s}^2}{5cd^2e_{\rm s}}\sim 3\times
10^{-4}\frac{E_{54}^{1/2}n_2^{1/2}L_{39.5}R_{17}^{1/2}} {e_{s,1}d_{28}^2} \quad {\rm cm^{-2}}, 
\ee
for the photon fluence for the case of on-axis star. 
Here, $d=10^{28}d_{28}$ cm is the proper distance of the burst that corresponds
to a redshift $z=1$ and  $L_{\rm s}=10^{39.5}L_{39.5}$ erg/s is the stellar luminosity.
The observed photon fluence depends weakly on the various parameters. In the
off-axis case the photon fluence is  a factor $\sim 1/\Gamma$ less. 

In deriving the last expression, I used that,
if the star is on-axis, the stellar radiation in the observer's cone 
is mainly located within a distance $\sim R_s/\Gamma$ from the O star.
The average energy density of the stellar photon field in this volume
is $\bar{U}_{\rm rad}=\int U_{\rm rad}dV/V=3L_{\rm s}\Gamma^2/4\pi c R_{\rm
s}^2$. The blast wave crosses this volume at a time $t_{cr}\sim 2 R_{\rm
s}/\Gamma c$. The total energy of the upscattered photons within the
observer's cone is $E_{\rm IC}^{\rm on} \sim 4c\sigma_{\rm T} \bar{U}_{\rm rad}
(4\gamma_{\rm e}\Gamma/3)^2 N t_{\rm cr}/3 \sim \Gamma \gamma_{\rm
 e}^2R_{\rm s}^2L_{\rm s}n\sigma_{\rm T}/c$. Dividing $E_{\rm IC}^{\rm on}$
with the typical energy of the upscattered photons $2\Gamma^2\gamma_{\rm
 e}^2e_{\rm s}$ and the surface area $\pi d^2/\Gamma^2$, one derives the
photon fluence on Earth (4). When  $1/\Gamma\ll \theta\ll 1$, the star 
is located at a distance $R_s\theta$ away from the line of sight. 
The energy density of stellar photons at this distance is 
$\bar{U}_{\rm rad}\simeq L_s/4\pi (R_s\theta)^2c$.
The blast wave crosses this volume at $t_{\rm cr} \sim 2R_s\theta/c$.  
When extrapolating the previous expressions for $\theta\sim 1$ (off-axis case), one finds 
$E_{\rm IC}^{\rm off}\sim E_{\rm IC}^{\rm on}/\Gamma$.

\subsubsection{Klein-Nishina suppression}

Equation~(4) holds as long as scattering takes place in the Thomson limit.
When the photon energy in the electron rest frame $E'\simeq 4/3\Gamma 
\gamma_e e_s$ exceeds the $m_{\rm e} c^2$,
relativistic effects decrease the scattering cross section 
and consequently the strength of inverse Compton emission.
For UV soft photons of $e_{\rm s}\sim 10$ eV, the Klein-Nishina suppression
in the cross section appears for observed $E_{IC}\simmore 25/(1+z)$ GeV. 
For IR photons of  $e_{\rm s}\sim 1$ eV
(expected from a red supergiant), the Klein-Nishina suppression
appears for $E_{IC}\simmore 250/(1+z)$ GeV. In the extreme
relativistic $E'\gg m_ec^2$ regime, the scattering cross section
is 
\be
\sigma=3\sigma_T(\ln 2x+1/2)/8x,
\ee
where $x=E'/m_ec^2$. The Klein-Nishina suppression is 
taken into account in the results that follow.

\subsection{Attenuation of the produced $\gamma$-rays}

The $\gamma$-rays can be attenuated by interacting with ambient radiation fields
creating electron-positron pairs at the location of production or on the way to Earth. 
I have verified that the produced $\gamma$-rays are not attenuated
significantly by interacting with the radiation field of the O star, but
I did not include attenuation of the signal from interaction with the 
extragalactic background light. This becomes significant for {\it observed} $\gamma$-rays
of energy $E_{\gamma}\simmore 100$ GeV for a burst that takes place at
$z\simmore 1$ (e.g. Blanch \& Martinez 2005).

\section{Results: detection prospects}

The resulting energy and strength of the inverse Compton emission are
summarized in Figs.~2 and 3. The emission appears in the $\sim 100$
GeV and $\sim 1$ GeV energy range for a star located at a distance $R_{s} \sim 10^{17}$ cm 
(hereafter referred to as ``close encounter'') and $\sim 3\times 10^{17}$ cm 
(distant encounter) from the center of the explosion, respectively.

\begin{figure}
\resizebox{\hsize}{!}{\includegraphics[angle=270]{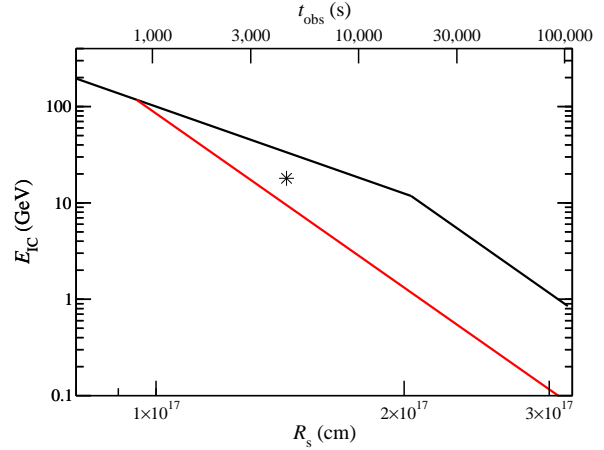}}
\caption[] { Observed energy of the scattered photons as a function of distance 
$R_s$ of the O star from the center of the explosion.
The black curve corresponds to a main-sequence 
O star and the red curve to a red supergiant respectively. $t_{\rm obs}$
stands for the time after the burst that the inverse Compton emission 
is observed. Photon energy and time are measured in the frame of the 
observer assuming a burst at $z=1$.
The asterisk stands for the 18 GeV photon detected by EGRET
$\sim 4500$ s after the burst of the 17 February 1994 assuming 
a redshift of $z=1$. 
\label{fig2}}
\end{figure}

The time after the $\gamma$-ray burst at which the observer sees the scattered photons 
depends on the distance $R_{\rm s}$ of the star (Sari et al. 1998)
\be
t_{\rm obs}\simeq \frac{R_s}{4\Gamma^2 c}(1+z)\simeq 520(1+z)\frac{n_2R_{17}^4}{E_{54}}\quad {\rm s}. 
\ee
The duration of the emission $\delta t$ is determined by the different travel
times of the scattered photons and is about $t_{\rm obs}$. 
A close encounter results in emission tens of  minutes
after the burst while a distant encounter is observed
about half a day after the burst.

The observed photon fluence from a burst at $z=1$ 
is shown in Fig. 3. It ranges from $\sim 5\times 10^{-7}$ photons/cm$^2$ 
to $\sim 5\times 10^{-3}$ photons/cm$^2$ for an off-axis main sequence O star 
and an on-axis red supergiant, respectively. 

A close encounter results in emission within the observed energy range of 
Cherenkov telescopes, especially that of MAGIC II and HESS II. With MAGIC's effective 
area of $\sim 3\times 10^8$ cm$^2$ at $\sim 100$ GeV (Albert et al. 2007),
this emission should be easily detectable with $\sim 10^2-10^5$ photon counts.
The emission from both close and distant encounters falls within the observed
energy range and, under favorable conditions, the sensitivity of the LAT 
telescope on the GLAST satellite. With a collective area of 
$\sim 10^4$ cm$^2$, LAT can detect $\sim 40$ photons from an on-axis
red supergiant. 

The GRB 940217 was one of the brightest ever detected.
The EGRET telescope detected several $\sim 100$ MeV photons and 
one 18 GeV photon $\sim$4500 sec after the burst (Hurley et al. 1994). The energy of this
mysterious late photon is compatible with an encounter of the blast wave with O-star radiation
at a distance of $\sim 1.5 \times 10^{17}$ cm (Fig. 2). The redshift of this burst is not known, 
but the implied fluence of high-energy afterglow photons,  of the order $\sim 2\times 10^{-3}$ 
cm$^{-2}$ in the $\sim 10$ GeV range, is compatible with this scenario for a redshift
of order unity. The detection of the scattered component in a burst of known
redshift can strongly constrain the properties of the companion star and of
the circumburst medium through Eqs. (3), (4), and (6).  
 
\begin{figure}
\resizebox{\hsize}{!}{\includegraphics[angle=270]{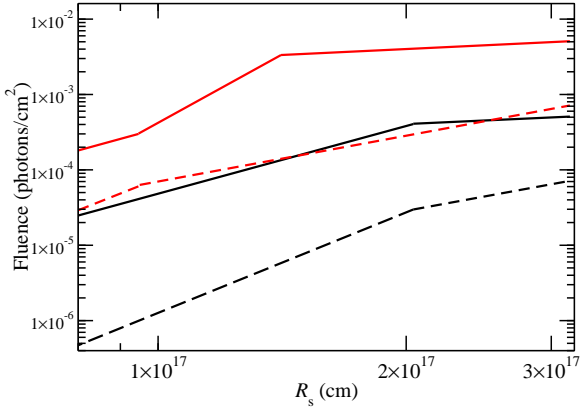}}
\caption[] {Observed fluence of scattered stellar photons
as a function of distance $R_s$ of the O star from the center of the 
explosion. The burst is assumed to take place at redshift $z=1$. 
The black curves correspond to a main-sequence O star and the red curves 
to a red supergiant. The solid curves correspond to the case 
where the star is located in the line of sight and the dashed curves to an off-axis star.
\label{fig3}}
\end{figure}

\section{Discussion and conclusions}

I studied the emission from the
interaction of the GRB blast wave with the dense photon field of an O star
located at $R_{\rm s} \sim 1-3\times 10^{17}$ cm from the center of the explosion.
Such distance is similar to the observed mean distance of O stars in the very dense 
stellar environments of massive clusters where a large fraction of the 
Wolf-Rayet stars reside.
   
The conclusion from this analysis is that  
inverse Compton scattering of the stellar photons by electrons accelerated in
the shock leads to powerful GeV emission that appears from just minutes up to one day
after the burst. This emission can be detected by Cherenkov 
telescopes and GLAST. Such detection can probe the properties of the blast
wave and the nearby star. Attenuation of the produced $\gamma$-rays by interaction 
with the extragalactic background light (EBL) is expected to result to a high-energy
cutoff of the $\gamma$-ray spectrum that will put constraints on the poorly known EBL.  

The external Compton emission discussed here appears at higher energy
than the  synchrotron-self-Compton emission predicted by the
standard afterglow model does. Assuming that a fraction $\epsilon_B\sim 10^{-3}$ 
of the energy released in the shock goes into
magnetic fields $B'$, the energy of the
synchrotron photons in the frame comoving with the flow 
$E'_{\rm syn}=3\gamma_{\rm e}^2eB'/4\pi m_ec\simeq 4 
\epsilon_{\rm e,-1}^2\epsilon_{\rm B,-3}^{1/2}E_{54}^{3/2}n_2^{-1}R_{17}^{-9/2}$ eV
is much lower than that of the stellar photons $E'_{\rm s}=\Gamma e_{\rm s}=400
E_{54}^{3/2}e_{\rm s,1} n_2^{-1}R_{17}^{-3/2}$ eV for the distances of interest. 

In this work, the simplification of a constant density circumburst medium is used,
although the medium, a product of the collision of stellar winds, is much more
complex. The wind interaction consists 
of regions of $r^{-2}$ density profiles close to the stars and regions of 
shocked winds (Stevens et al. 1992). A shock propagating in 
this environment may result in complex optical and X-ray afterglow
lightcurves (Ramirez-Ruiz et al. 2001; Nakar \& Granot 2007) 
that can also probe such an interaction.
 
The conditions discussed here are different from interaction
with photon fields closer in the center of explosion. These  include the photon field
in the progenitor's funnel that can seed the GRB emission 
(Lazzati et al. 2000) and the
shock-breakout photon field that can power an X-ray/$\gamma$-ray
precursor to the main burst (MacFadyen et al. 2001; Ramirez-Ruiz et  
al. 2002; Waxman \& Meszaros 2003). Furthermore, if present, the photon
field of a stellar companion, combined with prolonged central engine
activity, can result in afterglow $\gamma$-ray emission (Ramirez-Ruiz 2004).
These fields are located at distances $R\simless 10^{13}$ cm from the
center of the explosion, and they can limit the maximum Lorentz
factor to which the flow is accelerated through the so-called 
``Compton drag'' (e.g. Ghisellini et al. 2000).

\begin{acknowledgements}
I thank Henk Spruit for important suggestions and discussions and 
Jonathan Braithwaite for comments on the manuscript. 
\end{acknowledgements}


\begin{thebibliography}{}


\bibitem{}Albert, J. et al.\ 2007, \apj, 667, 358
\bibitem{} Blandford, R.~D., \& McKee, C.~F.\ 1976  Phys. Fluids, 19, 1130
\bibitem{} Blanch, O., \& Martinez, M.\ 2005,  Astropar. Phys.,
 23, 588 
\bibitem{} Chevalier, R.~A., \& Li, Z.-Y.\ 2000, \apj, 536,  195
\bibitem{} Crowther, P.~A. 2007, Annu. Rev. Astron. Astroph., 45, 177
\bibitem{} Figer, D.~F. 2004, in The Formation and Evolution of Massive
Young Star Clusters, ASP Conf. Ser., ed. H.~J.~G.~L.~M. Lamers, L.~J.~
Smith \& A. Nota (San Francisco: ASP), 322, 49
\bibitem{} Galama, T.~J., et al.\ 1998, Nature, 395, 670
\bibitem{} Ghisellini, G., Lazzati, D., Celotti, A., \& Rees, M.~J.\ 2000, \mnras, 316, L45 
\bibitem{} Hjorth, J., et al.\ 2003, Nature, 423, 847
\bibitem{} Hurley, K. et al.\ 1994, Nature, 372, 652
\bibitem{} Lazzati, D., Ghisellini, G., Celotti, A., \& Rees, M.~J.\ 2000, \apjl, 529, L17 
\bibitem{} MacFadyen, A.~I., \& Woosley, S.~E.\ 1999, \apj, 524, 262
\bibitem{} MacFadyen, A.~I., Woosley, S.~E., \& Heger, A.\ 2001, \apj, 550, 410 
\bibitem{} Massey, P., \& Hunter, D.~A.\ 1998, \apjl, 493,  180
\bibitem{} Nakar, E., \& Granot, J.\ 2007, \mnras, 380, 1744  
\bibitem{} Ramirez-Ruiz, E., Dray, L.~M., Madau, P., \& Tout, C.~A.\ 2001,
\mnras, 327, 829
\bibitem{} Ramirez-Ruiz, E., MacFadyen, A.~I., \& Lazzati, D.\ 2002, \mnras,
  331, 197 
\bibitem{} Ramirez-Ruiz, E.\ 2004, \mnras, 349, L38
\bibitem{} Sari, R., Piran, T., \& Narayan, R.\ 1998, \apjl, 497, 17
\bibitem{} Stanek, K.~Z., et al.\ 2003, \apjl, 591, 17
\bibitem{} Stevens, I.~R., Blondin, J.~M., \& Pollock, A.~M.~T.\ 1992,
\apj, 386, 265
\bibitem{} Waxman, E., \& M{\'e}sz{\'a}ros, P.\ 2003, \apj, 584, 390
\bibitem{} Wijers, R.~A.~M.~J., \& Galama, T.~J.\ 1999, \apj, 523, 177

\end{thebibliography}
\end{document}